\documentclass[12pt]{article}

\newcommand{\br}{{\bf R}}

\newcommand{\cov}{{\rm {cov}}}
\newcommand{\s}{{S_{\rm{symp}}}}
\newcommand{\cv}{{\cal V_{\rm{symp}}}}

\newcommand{\Om}{\Omega}

\usepackage{latexsym,amsfonts,amsmath,amsthm,amssymb,graphicx}

\usepackage[T1]{fontenc}
\usepackage[latin1]{inputenc}

\date{}
\title{Nonlinear Schr\"odinger equations from prequantum classical
statistical field theory}
\author{Andrei Khrennikov\\
International Center for
Mathematical Modeling \\
in Physics and Cognitive Sciences\\
University of V\"axj\"o, Sweden\\
Email:Andrei.Khrennikov@msi.vxu.se}

\begin{document}
\maketitle

 \abstract{We derive some important features of the standard quantum mechanics from
a certain classical-like model -- prequantum classical statistical
field theory, PCSFT. In this approach correspondence between
classical and quantum quantities is established through  asymptotic
expansions. PCSFT induces not only linear Schr\"odinger's equation,
but also its {\it nonlinear generalizations.}  This coupling with
``nonlinear wave mechanics'' is used to evaluate the small parameter
of PCSFT.}


\section{Introduction}

We demonstrated in [1] that averages of quantum observables can be
considered as approximations of averages of classical variables on
Hilbert phase space, see similarities but also differences with the
schemes of [2]--[4] (see also [5]-[12] on various attempts to go
beyond quantum mechanics). By choosing a special representation of
the infinite dimensional configuration space, namely as the
$L_2({\bf R}^3)$-space, we obtain the following representation of
the phase space:
\begin{equation}
\label{FFR} \Omega=L_2({\bf R}^3)\times L_2({\bf R}^3).
\end{equation}
In such a representation points of the prequantum phase space can be
interpreted as classical (vector) fields, $\psi(x)=(q(x), p(x)).$
Therefore we called our model {\it prequantum classical statistical
field theory,} PCSFT.

The correspondence between PCSFT and QM is {\it asymptotic}: we
expand classical  (``prequantum'') averages with respect to a small
parameter $\alpha$ -- the dispersion of {\it vacuum fluctuations.}
The main term in this expansion is given by the quantum average. We
call our approach asymptotic dequantization.  We emphasize that we
develop the classical (statistical) theory which {\it does not
reproduce exactly QM.} There is only an asymptotic coupling between
classical and quantum models. To come with concrete experimental
predictions, we should evaluate the small parameter $\alpha$ of the
model. We do this in the present note by exploring the relation
between PCSFT and theory of nonlinear Schr\"odinger's equation --
``nonlinear wave mechanics,'' [13]-[16].

 We point out that PCSFT induces not only conventional
linear Schr\"odinger's equation, but also so called {\it nonlinear
Schr\"odinger's equations,} cf. [13]-[16]. In PCSFT these nonlinear
Schr\"odinger equations appear very naturally as {\it infinite
dimensional Hamilton equations.} Hence the PCSFT-approach to QM
could be considered as additional justification of nonlinearity of
dynamics on prequantum scales. On the other hand, since the small
parameter $\alpha$ of PCSFT can be identified  with the coupling
constant (denoted by $b$ in [13], [14], [16]) in nonlinear
Schr\"odinger equations, we can apply experimental estimates [14]
for this coupling constant to evaluate the $\alpha,$ see section 7.

\section{Asymptotic statistical correspondence between classical and quantum models}

We define {\it ``classical statistical models''} in the following
way, see [1] for more detail: a) physical states $\omega$ are
represented by points of some set $\Omega$ (state space); b)
physical variables are represented by functions $f: \Omega \to {\bf
R}$ belonging to some functional space $V(\Omega);$ c) statistical
states are represented by probability measures on $\Omega$ belonging
to some class $S(\Omega);$ d) the average of a physical variable
(which is represented by a function $f \in V(\Omega))$ with respect
to a statistical state (which is represented by a probability
measure  $\rho \in S(\Omega))$ is given by
\begin{equation}
\label{AV0} < f >_\rho \equiv \int_\Omega f(\psi) d \rho(\psi) .
\end{equation}

A {\it classical statistical model} is a pair $M=(S, V).$ We recall
that classical statistical mechanics on the phase space
$\Omega_{2n}= {\bf R}^n\times {\bf R}^n$ gives an example of a
classical statistical model. But we shall not be interested in this
example in our further considerations. We shall develop  a classical
statistical model with {\it infinite-dimensional phase-space.}

The conventional quantum statistical model with the complex Hilbert
state space $\Omega_c$ is described in the following way (see
Dirac-von Neumann [17], [18] for the conventional complex model): a)
physical observables are represented by operators $A: \Omega_c \to
\Omega_c$ belonging to the class of continuous self-adjoint
operators ${\cal L}_s \equiv {\cal L}_s (\Omega_c);$ b) statistical
states are represented by von Neumann density operators, see [18]
(the class of such operators is denoted by ${\cal D} \equiv {\cal D}
(\Omega_c));$ d) the average of a physical observable (which is
represented by the operator $A \in {\cal L}_s (\Omega_c))$ with
respect to a statistical state (which is represented
  by the density operator $D \in {\cal D} (\Omega_c))$ is given by von Neumann's
formula [18]:
\begin{equation}
\label{AV1} <A >_D \equiv \rm{Tr}\; DA
\end{equation}
The {\it quantum statistical model} is the pair $N_{\rm{quant}}
=({\cal D}, {\cal L}_s).$

We are looking for a classical statistical model $M=(S, V)$ which
will give ``dequantization'' of the quantum model $N_{\rm{quant}}
=({\cal D}, {\cal L}_s).$ Here the meaning of ``dequantization''
should be specified. In fact, all ``NO-GO'' theorems (e.g., von
Neumann, Kochen-Specker, Bell,...) can be interpreted as theorems
about impossibility of various dequantization procedures. Therefore
we should define the procedure of dequantization in such a way that
there will be no contradiction with known ``NO-GO'' theorems, but
our dequantization procedure still will be natural from the physical
viewpoint. We define (asymptotic) dequantization as a family
$M^\alpha=(S^\alpha, V)$ of classical statistical models depending
on a small parameter $\alpha \geq 0.$ There  should exist maps
$T:S^\alpha\to {\cal D}$ and $T: V \to {\cal L}_s$ such that: a)
both maps are {\it surjections} (so all quantum states and
observables can be represented as images of classical statistical
states and variables, respectively); b) the map $T: V \to {\cal
L}_s$ is ${\bf R}$-linear (we recall that we consider real-valued
classical physical variables); c) classical and quantum averages are
coupled through the following asymptotic equality:
 \begin{equation}
\label{AQ} < f >_\rho =\frac{\alpha}{2}  <T(f)>_{T(\rho)} +
o(\alpha), \; \; \alpha \to 0
\end{equation}
(here $<T(f)>_{T(\rho)}$ is the  quantum average). This equality can
be interpreted in the following way. Let $f(\psi)$ be a classical
physical variable (describing properties of microsystems - classical
fields having very small magnitude $\alpha).$  We define its {\it
amplification} by:
$f_\alpha (\psi) =\frac{1}{\alpha} f(\psi).$
 If we interpret $\alpha$ as the intensity of vacuum fluctuations,
then $f_\alpha (\psi)$ is the {\it relative intensity of $f(\psi)$
with respect to vacuum fluctuations.} By dividing both sides of the
equation (\ref{AQ}) by $\alpha$ we obtain:
\begin{equation} \label{AQ4}
< f_\alpha >_\rho = \frac{1}{2}<T(f)>_{T(\rho)} + o(1), \; \; \alpha
\to 0,
\end{equation}
Hence: {\it The quantum term gives the main contribution into the
relative intensity with respect to vacuum fluctuations.} The crucial
difference from dequantizations considered in known ``NO-GO''
theorems (e.g., von Neumann, Kochen-Specker, Bell) is that in our
case {\it classical and quantum averages are equal only
asymptotically.}

\section{Prequantum classical statistical model with the infinite dimensional phase space}

We choose the phase space $\Om= Q\times P,$ where $Q=P=H$ and $H$ is
the infinite-dimensional real (separable) Hilbert space. We consider
$\Omega$ as the real Hilbert space with the scalar product $(\psi_1,
\psi_2)= (q_1, q_2) + (p_1, p_2).$ We denote  by $J$ the symplectic
operator on $\Omega:
 J= \left( \begin{array}{ll}
 0&1\\
 -1&0
 \end{array}
 \right ).$
Let us consider the class ${\cal L}_{\rm symp} (\Omega)$ of bounded
${\bf R}$-linear operators $A: \Omega \to \Omega$ which commute with
the symplectic operator:
\begin{equation}
\label{SS} A J= J A
\end{equation}
This is a subalgebra of the algebra of bounded linear operators
${\cal L} (\Omega).$ We also consider the space of ${\cal
L}_{\rm{symp}, s}(\Omega)$ consisting of self-adjoint operators.

By using the operator $J$ we can introduce on the phase space
$\Omega$ the complex structure. Here $J$ is realized as $-i.$ We
denote $\Omega$ endowed with this complex structure by $\Omega_c:
\Omega_c\equiv Q\oplus i P.$ We shall use it later. At the moment
consider $\Omega$ as a real linear space and consider its
complexification $\Omega^{{\bf C}}= \Omega \oplus i \Omega.$

Let us consider the functional space ${\cal V}_{\rm{symp}}(\Omega)$
consisting of functions $f:\Omega \to {\bf R}$ such that: a) the
state of vacuum is  preserved : $f(0)=0;$ b) $f$ is $J$-invariant:
$f(J\psi)= f(\psi);$ c) $f$ can be extended to the  analytic
function $f:\Omega^{{\bf C}}\to {\bf C}$ having  the exponential
growth: $ \vert f(\psi)\vert \leq c_f e^{r_f \Vert \psi \Vert}$ for
some $c_f, r_f \geq 0$ and for all $\psi\in \Omega^{{\bf C}}.$ We
remark that the possibility to extend a function $f$ analytically
onto $\Omega^{{\bf C}}$ and the exponential estimate on
$\Omega^{{\bf C}}$ plays the important role in the asymptotic
expansion of integrals over the infinite dimensional phase
space.\footnote{To get a mathematically rigor formulation,
conditions in [1] should be reformulated in the similar way.} But
this is purely mathematical condition which could be in principle
weakened.

The following elemntary mathematical result plays the fundamental
role in establishing classical $\to$ quantum correspondence: {\it
Let $f$ be a smooth $J$-invariant function. Then } $f^{\prime
\prime}(0)\in {\cal L}_{\rm{symp}, s}(\Omega).$ In particular, a
quadratic form is $J$-invariant iff it is determined by an operator
belonging to ${\cal L}_{\rm{symp}, s}(\Omega).$

We consider the space statistical states
$S_{\rm{symp}}^{\alpha}(\Omega)$ consisting of measures $\rho$ on
$\Omega$ such that: a) $\rho$ is symmetric (in particular, its mean
value equals to zero); b) the dispersion of $\rho$ is equal to
$\alpha:$
$$
\sigma^2(\rho)= \int_\Omega \Vert \psi\Vert^2 d \rho(\psi)= \alpha,
\; \alpha \to 0;
$$
c) $\rho$ is $J$-invariant;  d) for any $r>0$ the exponential
function $e^{r \Vert \psi \Vert}$ is integrable.  The latter
condition is a purely mathematical one and it could be weakened.
Such measures describe small statistical fluctuations of the vacuum
field.

 The following trivial mathematical result plays the
fundamental role in establishing classical $\to$ quantum
correspondence: {\it Let a measure $\rho$ be $J$-invariant. Then its
covariation operator} $B= \rm{cov}\; \rho \in {\cal L}_{\rm{symp},
s}(\Omega).$ Here $(By_1, y_2)= \int (y_1, \psi)(y_2, \psi) d \rho(
\psi).$

We now consider the complex realization $\Omega_c$ of the phase
space and the corresponding complex scalar product $<\cdot, \cdot>.$
We remark that the class of operators ${\cal L}_{\rm symp} (\Omega)$
is mapped onto the class of ${\bf C}$-linear operators ${\cal
L}(\Omega_c).$

We also define for any measure its complex covariation operator
$B^c= \rm{cov}^c \rho$ by $ <B^c y_1, y_2>=\int <y_1, \psi> <\psi,
y_2> d \rho (\psi). $ We remark that for a $J$-invariant measure
$\rho$ its complex and real covariation operators are related as
$B^c=2 B.$

As in the real case [1], we  can prove that for any operator $ A\in
{\cal L}_{\rm{symp}, s}(\Omega):$
\begin{equation}
\label{CI2} \int_\Omega <A\psi,\psi> d \rho (\psi) = \rm{Tr}
\;\rm{cov}^c \rho \;A.
\end{equation}
We pay attention that the trace is considered with respect to the
complex inner product. We remark that this formula has been already
used by Bach [19] for representing quantum averages. The crucial
difference from Bach's approach is that we consider arbitrary
functionals of $\psi$ and we shall obtain an {\it asymptotic
relation between classical and quantum averages.}

We consider now the one parameter family of classical statistical
models:
\begin{equation}
\label{MH} M^\alpha= ( S_{\rm{symp}}^\alpha(\Omega),{\cal
V}_{\rm{symp}}(\Omega)), \; \alpha\geq 0,
\end{equation}

{\bf Theorem 1.} {\it Let $f \in {\cal V}_{\rm{symp}}(\Omega)$ and
let $\rho \in S_{\rm{symp}}^\alpha(\Omega).$ Then the following
asymptotic equality holds:
\begin{equation}
\label{ANN3} <f>_\rho =  \frac{\alpha}{2} \; \rm{Tr}\; D^c \;
f^{\prime \prime}(0) + o(\alpha), \; \alpha \to 0,
\end{equation}
where the operator $D^c= \rm{cov}^c \; \rho_{\rm{scal}}$ and $
\rho_{\rm{scal}}$ is the $\sqrt{\alpha}$-scaling of the measure
$\rho.$ Here
$o(\alpha) = \alpha^2 R(\alpha, f, \rho),$
where $\vert R(\alpha,f,\rho)\vert \leq c_f\int_\Omega  e^{r_f \Vert
\Psi \Vert}d \rho_{\rm{scal}} (\Psi).$ }

Here and everywhere below we use the symbols $\psi$ and $\Psi$ to
denote ``prequantum field'' and its scaling, respectively:
\begin{equation}
\label{OLV} \psi=\sqrt{\alpha} \Psi.
\end{equation}
We see that the classical average (computed in the model $M^\alpha=
( S_{\rm{symp}}^\alpha(\Omega),{\cal V}_{\rm{symp}}(\Omega))$ by
using the measure-theoretic approach) is coupled through
(\ref{ANN3}) to the quantum average (computed in the model
$N_{\rm{quant}} =({\cal D}(\Omega_c),$ ${\cal L}_{{\rm
s}}(\Omega_c))$ by the von Neumann trace-formula).

The equality (\ref{ANN3}) can be used as the motivation for defining
the following classical $\to$ quantum map $T$ from the classical
statistical model $M^\alpha= ( S_{G, \rm{symp}}^\alpha,{\cal
V}_{\rm{symp}})$ onto the quantum statistical model
$N_{\rm{quant}}=({\cal D}, {\cal L}_{{\rm s}}):$
\begin{equation}
\label{Q20} T: S_{\rm{symp}}^\alpha(\Omega) \to {\cal D}(\Omega_c),
\; \; D^c=T(\rho)= \rm{cov}^c\; \rho_{\rm{scal}}
\end{equation}
(the measure $\rho$ is represented by the density matrix $D^c$ which
is equal to the complex covariation operator of its
$\sqrt{\alpha}$-scaling );
\begin{equation}
\label{Q30} T: {\cal V}_{\rm{symp}}(\Omega) \to {\cal L}_{{\rm
s}}(\Omega_c), \; \; A_{\rm quant}= T(f)= f^{\prime\prime}(0).
\end{equation}
Our previous considerations can be presented as

\medskip

{\bf Theorem 2.} {\it The one parametric family of classical
statistical models $M^\alpha= (S_{\rm{symp}}^\alpha(\Omega),{\cal
V}_{\rm{symp}}(\Omega))$ provides asymptotic dequantization of the
quantum model $N_{\rm{quant}} =({\cal D}(\Omega_c),$ ${\cal L}_{{\rm
s}}(\Omega_c))$ through the pair of maps (\ref{Q20}) and
(\ref{Q30}). The classical and quantum averages are coupled by the
asymptotic equality (\ref{AQ4}).}

\section{Infinite dimension of physical space}

This is a good place to discuss the model of physical space in
PCSFT. Here the {\it real physical space} is Hilbert space. If we
choose the realization $H=L_2({\bf R}^3),$ then we obtain the
realization of $H$ as the space of classical fields on ${\bf R}^3.$
So the {\it conventional space ${\bf R}^3$ appears only through this
special representation of the Hilbert configuration space.} Dynamics
in ${\bf R}^3$ in just a shadow of dynamics in the space of fields.
However, we can choose other representations of the Hilbert
configuration space. In this way we shall obtain classical fields
defined on other ``physical spaces.''

We remark that at the first sight the situation with development of
PCSFT is somewhat reminiscent of the one confronted by Scr\"odinger
in his introduction of his wave equation, which ``maps'' waves in
the configuration space. However,  even though he had, just as did
Einstein, major reservations concerning quantum mechanics as the
ultimate theory of quantum phenomena, Scr\"odinger never went so far
as to see any space other than  ${\bf R}^3$ as real.\footnote{We
point out L. De Broglie in his theory of double solution  emphasized
the fundamental role of physical space ${\bf R}^3.$ Such a viewpoint
also was  common for adherents of Bohmian mechanics (in any case for
D. Bohm and J. Bell). However, {\it recently B. Hiley started to
consider the momentum representation of Bohmian mechanics} [20] and
it seems that in Hiley's approach to Bohmian mechanics the position
representation does not play an exceptional role.}

On the other hand, string theory does introduce spaces of higher
dimensions, although not of infinite dimensions. This approach was
one of inspirations for our radical viewpoint to physical space. One
could speculate that on scales of quantum gravity and string theory
space became infinite dimensional, just as those theories the space
has the (finite) dimension higher than three.\footnote{In our
approach quantum theory is not the ultimate theory. It has its
boundaries of applications. Therefore there are no reasons to expect
that ``quantum gravity'' should exist at all. Thus it would be
better to speak not about scales of quantum gravity and string
theory, but simply about the Planck scale for length and time.}

\section{Hamilton-Schr\"odinger dynamics}

States of systems with the infinite number of degrees of freedom -
classical fields -- are represented by points $\psi=(q, p) \in \Om;$
evolution of a state is described by the Hamiltonian equations:
\begin{equation}
\label{HE} \dot q = \frac{\partial {\cal H}}{\partial p},\; \;  \dot
p=-\frac{\partial {\cal H}}{\partial q}.
\end{equation}
First we consider a quadratic Hamilton function: ${\cal H}(q,
p)=\frac{1}{2} ({\bf H} \psi,\psi),$ where ${\bf H}: \Om \to \Om$ is
an arbitrary symmetric (bounded) operator. In this special case the
Hamiltonian equations have the form: $\dot q= {\bf H}_{21}q + {\bf
H}_{22} p, \; \; \dot p=-( {\bf H}_{11}q +{\bf H}_{12}p),$ or
$\dot \psi= \left( \begin{array}{ll}
\dot q\\
\dot p
\end{array}
\right )=J{\bf H} \psi.$
Thus {\it quadratic Hamilton functions induce linear Hamilton
equations.} We get $\psi(t)= U_t \psi, \; \; \mbox{where} \;
U_t=e^{J {\bf H} t}.$ The map $U_t\psi$ is a linear Hamiltonian flow
on the phase space $\Omega.$

Let us  now consider an operator ${\bf H} \in {\cal L}_{\rm symp, s}
(\Omega)$: ${\bf H}= \left(
\begin{array}{ll}
R&T\\
-T&R
\end{array}
\right).$ This operator defines the quadratic  Hamilton function $
{\cal H}(q, p)=$\\ $\frac{1}{2}[(R p, p) + 2 (Tp, q) + (Rq, q)], $
where $R^*=R , \; \; T^*=-T.$ Corresponding Hamiltonian equations
have the form $\dot q=Rp-Tq, \;  \dot p=-(Rq + Tp).$ We pay
attention that for a  $J$-invariant Hamilton function, the
Hamiltonian flow $U_t \in {\cal L}_{\rm{symp}}(\Omega).$ By
considering the complex structure on the infinite-dimensional phase
space $\Omega$ we write the latter Hamiltonian equations in the form
of the Sch\"odinger equation on $\Omega_c:$
$$i  \frac{d \psi}{d t} = {\bf H} \psi;$$
its solution has the following complex representation: $\psi(t)=U_t
\psi, \; \; U_t=e^{-i{\bf H} t}.$ We consider the Planck system of
units in that $h=1.$ This is {\it the complex representation of
flows corresponding to quadratic $J$-invariant Hamilton functions.}

\medskip

However, in our approach there are no reasons to restrict
considerations by quadratic ($J$-invariant) Hamilton functions. We
can take any function ${\cal H}\in C^1(\Omega)$ as the Hamilton
function. It induces the Hamilton-Schr\"odinger dynamics:
\begin{equation}
\label{HEX} i \frac{d \psi}{d t} = {\cal H}^\prime(\psi).
\end{equation}
We pay attention that the Hamilton-Schr\"odinger equation can be
considered for any Hamilton function which is one time continuously
Frechet-differentiable on the infinite dimensional phase space.
However, we obtained the asymptotic expansion of classical average
(on the infinite dimensional phase space) only for analytic physical
variables. Our approach could be generalized for functions of the
class $C^2.$ But if a function does not have the second
Frechet-derivative, then we are not able to apply technique based on
the Taylor expansion and obtain quantum effects (which give the
contribution of the second order in the Taylor formula), cf.
problems which will be discussed in section 6.

\section{Scaling of prequantum variables}
Considerations of section 3, as well as everywhere in [1], were
based on scaling of probability measures representing classical
statistical states. Since there is the natural duality between
measures and functions, it is possible to consider scaling of
classical variables, instead of scaling of measures. It is important
to consider scaling of variables, because the basic equation of
quantum mechanics is Schr\"odinger's equation (which is an image of
the classical Hamiltonian equation). For any function $f: \Omega \to
\br$ we set
\begin{equation}
\label{SN} f_Q(\Psi)=\frac{1}{\alpha} f(\sqrt{\alpha} \Psi).
\end{equation}
The $f_Q$ is the result of the transition from the prequantum system
of the $\psi$-field coordinates to the quantum system of the
$\Psi$-field coordinates and $1/\alpha$-amplification of the
classical variable $f$. Such a renormalization of $f$ can be
justified in two ways: a) through statistical averages, b) through
the Hamiltonian-Schr\"odinger equations.

\subsection{Statistical origin of the renormalization}
The asymptotic representation of the classical average can be
written as
$$
<f/\alpha>_{\rho}=
 \int_\Omega \frac{f(\sqrt{\alpha}\Psi)}{\alpha} d\rho_{\rm{scal}}(\Psi) \equiv
\int_\Omega f_Q (\Psi) d\rho_{\rm{scal}} (\Psi)
$$
$$
= \frac{1}{2}\rm{Tr} \; \cov^c \rho_{\rm{scal}}
\;f^{\prime\prime}(0) + o(1), \alpha \to 0.
$$
We remark that
$\frac{\partial^2 f}{\partial \psi^2}(0)=\frac{\partial^2
f_Q}{\partial \Psi^2}(0).$
Thus we have:
$<f_Q>_{\rho_{\rm{scal}}}=\frac{1}{2} \rm{Tr} \; \cov^c
\rho_{\rm{scal}} \; f_Q^{\prime \prime}(0) + o(1), \alpha \to 0.$
Therefore we can consider a new classical model in that statistical
states are given by measures $\mu \in \s (\Omega)$  (having
dispersion 1) and physical variables by functions $g \in \cv
(\Omega)$ (we note that any function $g$ can be considered as
scaling  for some $f$ belonging the same functional space:
$g(\Psi)\equiv f_Q(\Psi)=f(\sqrt{\alpha}\Psi)/\alpha$).\footnote{In
other words this functional space is invariant with respect to the
scaling map: $f \to f_Q.$} We consider the classical statistical
model \\ $M =(\s(\Omega), {\cal V}_{\rm{symp}}(\Omega))$ and the map
$T: M \to N_{\rm{quant}},$ given by:
\begin{equation}
\label{LE} T(\mu)=\cov^c \mu
\end{equation}
\begin{equation}
\label{LE1} T(g)=g^{\prime\prime}(0)
\end{equation}
This map provides dequantization of $\bf N$ and the following
asymptotic equality holds:
\begin{equation}
\label{LE2} <f_Q>_\mu=\frac{1}{2} <T(f_Q)>_{T(\mu)} + o(1), \alpha
\to 0.
\end{equation}

\subsection{Dynamical origin of renormalization}
Let us consider a Hamilton function ${\cal H}(\psi).$ The
corresponding Hamilton equation is given by (\ref{HEX}). We now
change the system of coordinates: $\psi=\sqrt{\alpha} \Psi$ and
obtain the Hamilton equation in the $\Psi$-coordinates:
\begin{equation}
\label{LE4}  i \dot \Psi= {\cal H}_Q^\prime(\Psi).
\end{equation}
If ${\cal H}(\psi)$ is a quadratic (J-invariant) function, then
${\cal H}(\psi)\equiv {\cal H}_Q (\Psi)$ and we obtain the
conventional Schr\"odinger equation. Thus we came again to the same
procedure of renormalization ${\cal H} \to {\cal H}_Q$ of classical
physical variables. If ${\cal H}$ is not quadratic, then the
Schr\"odinger-Hamilton equation (\ref{LE4}) gives, in particular, so
called nonlinear Schr\"odinger equations, see, e.g., [13]--[16]. For
example, let ${\cal H}(\psi)=\frac{1}{2} (\hat H \psi, \psi) +
\frac{1}{4} {\cal H}_4 (\psi, \psi, \psi, \psi),$ where ${\cal H}_4$
is a $J$-invariant form of degree four. Then
$${\cal H}_Q (\Psi)=\frac{1}{2} (\hat H \Psi, \Psi) +
\frac{\alpha}{4} {\cal H}_4 (\Psi, \Psi, \Psi, \Psi).$$ Hence, the
corresponding Schr\"odinger-Hamilton equation has the form:
\begin{equation}
\label{OL} i \dot \Psi= \hat H \Psi + \alpha {\cal H}_4^\prime
(\Psi, \Psi, \Psi).
\end{equation}
For example, let $ \Omega= L_2^c (\br^3)$ and let $${\cal H}_4
(\Psi, \Psi, \Psi, \Psi)= \int_{\br^{12}}K_4 (x_1, x_2, x_3, x_4)
\Psi (x_1) \Psi (x_2) \overline{\Psi(x_3)}\; \overline{\Psi(x_4)} d
x_1^3 \ldots dx_4^3.$$ By choosing the kernel $k(x_1, x_2, x_3,
x_4)= \delta(x_1-x_2) \delta(x_2-x_3) \delta(x_3-x_4),$ we obtain
the Hamilton function with nonquadratic term: ${\cal H}_4(\Psi,
\Psi, \Psi, \Psi)= \int_{\br^3}|\Psi(x)|^4 dx,$ and the well known
nonlinear Schr\"odinger's equation
\begin{equation}
\label{ENE} i \frac{\partial \Psi}{\partial t} (t, x)= -\frac{1}{2}
\Delta \Psi (t, x) + V(x) \Psi(t, x) + \alpha |\Psi(t, x)|^2 \Psi
(t, x).
\end{equation}
However, PCSFT induces the class of Schr\"odinger-Hamilton equations
which is essentially larger than the conventional class of nonlinear
Schr\"odinger's equations with polynomial nonlinearities. For
example, let us choose ${\cal H}_4 (\Psi, \Psi, \Psi, \Psi)=
\frac{1}{4} (\Gamma_1 \Psi, \Psi) (\Gamma_2 \Psi, \Psi),$ where
$\Gamma_j: \Omega \to \Omega$ are $J$-invariant linear operators. We
obtain the Schr\"odinger-Hamilton equation:
\begin{equation}
\label{OL1} i \frac{\partial \Psi}{\partial t}= \hat H \Psi +
\frac{\alpha}{2} [(\Gamma_1 \Psi, \Psi) \Gamma_2 \Psi + (\Gamma_2
\Psi, \Psi) \Gamma_1 \Psi]
\end{equation}
On the other hand, in conventional nonlinear wave mechanics there
were considered equations corresponding to Hamilton functions ${\cal
H} (\psi)$ which are not analytic (and even not twice
differentiable). The most important example are log-nonlinearities,
see [13]-[16]. We shall come back to this question in section 7.

\section{An estimation of the small parameter $\alpha$ through theory of
nonlinear Schr\"odinger equations}

The derivation of nonlinear Schr\"odinger equations in the standard
Hamiltonian formalism (but on the infinite dimensional phase space)
is an important consequence of PCSFT. Nonlinear Schr\"odinger's
equations were considered in many papers, see, e.g., [13]--[16]. And
the problem of experimental verification was already studied [14],
[16]. In PCSFT the parameter $\alpha$ can also be interpreted as the
coupling constant for nonlinear perturbations of the Schr\"odinger
equation. Hence, the upper bound for such a constant that was
obtained in theory of nonlinear Schr\"odinger equations [14], [16]
is also valid for the small parameter $\alpha$ of PCSFT.

The main problem in applications of results  obtained in [13]--[16]
for evaluating the small parameter $\alpha$ of PCSFT is that
research in the conventional nonlinear wave mechanics was mainly
concentrated on the equation with {\it log-nonlinearity.} I know
only one paper on experimental estimation of the coupling constant
for nonlog-nonlinearities, see [14] (Weinberg), but even in this
paper finally there was considered the log-case. We recall that by
reasons of locality the following equation plays the fundamental
role in conventional nonlinear mechanics, see [15]:
\begin{equation}
\label{NL} i \frac{\partial \Psi}{\partial t}= - \frac{1}{2} \Delta
\Psi + V \Psi + b \ln |\Psi|^2 \Psi
\end{equation}
There was obtained the experimental estimate of the coupling
constant $b,$ see [14], [16]:
\begin{equation}
\label{EE} |b| \leq 10^{-15} e V.
\end{equation}
We pay attention that (\ref{NL}) is the Hamilton equation on
$\Omega$ with the Hamilton function
\begin{equation}
\label{GM1} {\cal H}_Q (\Psi)= \frac{1}{2} \int_{\br^3}
\Big[\frac{|\nabla \Psi(x)|^2}{2} + b|\Psi(x)|^2
(\ln|\Psi(x)|^2-1)\Big] d^3x.
\end{equation}
The main problem is that this function is not of the $C^2$-class on
the phase space $\Omega=L_2 (\br^3) \times L_2 (\br^3).$ Therefore
we could not use the Taylor expansion and apply directly our scheme
of asymptotic dequantization. Nevertheless, let us try to proceed.

We now should be  careful with dimensions of quantities under
consideration. By our interpretation [1] of the background random
field $\psi(x),$ the $|\psi(x)|^2$ has the dimension of the density
of energy: $|\psi(x)|^2 \sim  E/L^3.$ Since for the conventional
$\Psi$-function, the $|\Psi(x)|^2$ has the dimension $\; \sim \;
1/L^3,$ we have that $\alpha$ {\it has the dimension of energy.} Let
us write a prequantum Hamilton function inducing the
log-nonlinearity by taking into account dimensions of quantities. We
start by repeating previous considerations for quantities with
physical dimensions. First we consider dynamics of the background
$\psi$-field. The Hamilton equation should be, in fact, written in
the form
$i \tau \frac{\partial \psi}{\partial t}= {\cal H}^\prime (\psi),$
where $\tau$ has the {\it dimension of time.} We choose $\tau=t_{
P}$ the {\it Planck time.} By setting now
${\cal H}_Q(\Psi)=\Big(\frac{E_{P}}{\alpha}\Big) {\cal
H}(\sqrt{\alpha} \Psi),$
we can write our equation in the $\Psi$-system of coordinates
\begin{equation}
\label{JB1} ih \frac{\partial \Psi}{\partial t}= {\cal
H}_Q^\prime(\Psi)
\end{equation}
For example, we know that for a nonrelativistic quantum particle in
the QM-approximation:
\begin{equation}
\label{JB2} {\cal H}_Q (\Psi)=\int_{\br^3} \Big[\frac{h^2}{2m}
|\nabla \Psi (x)|^2 + V (x)|\Psi (x)|^2\Big] d^3 x
\end{equation}
Therefore the corresponding prequantum Hamilton function:
\begin{equation}
\label{JB3} {\cal H}(\psi)=\int_{\br^3} \Big[\frac{h^2}{2m E_P}|
\nabla \psi(x)|^2 + \frac{V(x)}{E_P}|\psi(x)|^2\Big] d^3x.
\end{equation}
Here we have used the relation: ${\cal
H}(\psi)=\Big(\frac{\alpha}{E_P}\Big){\cal H}_Q
(\frac{\psi}{\sqrt{\alpha}}).$ In the real physical dimensions the
log-term of ${\cal H}_Q(\Psi),$ see (\ref{GM1}), should be written
as
\begin{equation}
\label{GM2} U_Q(\Psi)=b \int_{\br^3} |\Psi (x)|^2 (\ln a^3|\Psi
(x)|^2-1) d^3 x,
\end{equation}
where $a \sim  L$ (since $|\Psi (x)|^2 \sim  1/L^3).$ Its prequantum
counterpart
\begin{equation}
\label{GM3} U(\psi)= \Big(\frac{\alpha}{E_{P}}\Big)
\Big(\frac{b}{\alpha}\Big) \int_{\br^3} |\psi (x)|^2 (\ln
\frac{a^3}{\alpha}|\psi (x)|^2-1) d^3 x.
\end{equation}
It is natural to identify the energy $b$-scale corresponding to the
nonlinear perturbation of the conventional Schr\"odinger equation
and the $\alpha$-scale. In this case
\begin{equation}
\label{GM5} b=\alpha
\end{equation}
and\begin{equation} \label{GM6} U_Q(\Psi)=\alpha \int_{\br^3} |\Psi
(x)|^2 (\ln a^3|\Psi (x)|^2-1) d^3 x.
\end{equation}
Let $\mu \in \s (\Omega).$ Then $$<{\cal H}_Q>_\mu =\frac{1}{2}
\rm{Tr} \; \cov^c \mu \Big[- \frac{h^2}{2m} \Delta + V\Big]
$$
$$
+ \alpha \int_\Omega \Big(\int_{\br^3}|\Psi (x)|^2 (\ln
a^3|(x)|^2-1) d^3 x\Big) d\mu (\Psi)=\frac{1}{2} \rm{Tr} \; \cov^c
\mu \;{\bf H} + o(1),$$ $\alpha \to 0.$ Here ${\bf H}$ is the
quantum Hamiltonian. This indirect estimate of $\alpha$ gives us:
$|\alpha| \leq 10^{-15} e V.$ Of course, this is only the upper
bound and it might be that $\alpha$ should be, in fact, essentially
less than $10^{-15} e V.$

\section{Hamilton function for conventional nonlinear Schr\"odinger equations}

In the conventional nonlinear quantum mechanics, see [13], [15],
there are considered equations of the form:
\begin{equation}
\label{z1} ih \frac{\partial \psi}{\partial t}= \Big[-
\frac{h^2}{2m} \Delta + V + F(|\psi|^2)\Big] \psi
\end{equation}
In fact, the Hamilton function for this equation was given already
in [13] (but without interpreting the equation (\ref{z1}) as the
Hamilton equation in the infinite-dimensional phase space).
Following [] we set $U_Q (\Psi)=\frac{1}{2} \int_{\br^3} d^3 x
(\int_0^{|\Psi(x)|^2} dq F(q)).$ Then
$U_Q^\prime(\Psi)=F(|\Psi(x)|^2)\Psi(x).$ We pay attention that
$U_Q(\Psi)$ is $J$-invariant. To apply our theory, we need at least
the $C^2$-class for $U_Q: \Omega \to \Omega,$ thus at least
$C^1$-class for $F:\br \to \br.$

\bigskip

Results of this paper were presented in a series of author's talks
on PCSFT -- at Steklov Mathematical Institute of Russian Academy of
Science (sections of mathematical and theoretical physics as well as
the general institute's seminar), at Moscow Institute of Physical
Engineering, at University of Mannheim (department of mathematics),
at University of Bonn (department of stochastics), at Humboldt
University of Berlin (quantum optics laboratory). I would like to
thank S. Albeverio, O. Benson, E. Binz, A. Ezhov, A. Slavnov, V.
Vladimirov and I. Volovich for hospitality.

{\bf REFERENCES}

1. A. Yu. Khrennikov, Phys. A: Math. Gen. 38 (2005) 9051; Found.
Phys. Letters 18 (2005) 637.

2. B. Mielnik, Commun. Math. Phys. 31 (1974) 221.

3. R. Haag and U. Bannier, Commun. Math. Phys. 60 (1978) 1.

4. J. Bell and B. Hallet, Philos. Sci. 49 (1982) 355.

5. L. de la Pena and A. M. Cetto,  The Quantum Dice: An Introduction
to Stochastic Electrodynamics, Kluwer, Dordrecht, 1996; T. H. Boyer,
A Brief Survey of Stochastic Electrodynamics, in Foundations of
Radiation Theory and Quantum Electrodynamics,  A. O. Barut, ed.,
Plenum, New York, 1980.

6. D. Bohm and B. Hiley, The undivided universe: an ontological
interpretation of quantum mechanics, Routledge and Kegan Paul,
London, 1993.

7. P. Holland, The quantum theory of motion, Cambridge University
press, Cambridge, 1993.

8. E. Nelson , Quantum fluctuation, Princeton Univ. Press,
Princeton, 1985.

9. M. Davidson,  J.Math. Phys. 20  (1979) 1865; Physica A  96 (1979)
465.

10. P. Morgan, Phys. Lett A 338 (2005) 8; 321 (2004) 216.

11. G. `t Hooft,  hep-th/0105105; quant-ph/0212095.

12. W. M. De Muynck, Foundations of quantum mechanics, an
empiricists approach, Kluwer, Dordrecht, 2002.

13. I. Bialynicki-Birula, J. Mycielski,  Annals of Physics 100
(1976) 62.

14. A. Shimony, Phys. Rev. A 20  (1979) 394.

C. G. Shull, D. K. Atwood, J. Arthur, and M. A. Horne, Phys. Rev.
Letters 44 (1980) 765;

S. Weinberg, Phys. Rev. Letters 62  (1989) 485.

15. H. D. Doebner,  Non-linear partial differential operators and
quantization, Berlin, Springer-Verlag, 1983.

16. N. Gisin, Hel. Physica Acta 62 (1989) 363.

17. P. A. M.  Dirac, The Principles of Quantum Mechanics, Oxford
Univ. Press, 1930.

18. J. von Neumann, Mathematical foundations of quantum mechanics,
Princeton Univ. Press, Princeton, N.J., 1955.

19. A. Bach,  J. Math. Phys. 14 (1981) 125; Phys. Lett. A 73 (1979)
287; J. Math. Phys. 21 (1980) 789.

20. B. J. Hiley, From the Heisenberg Picture to Bohm: a New
Perspective on Active Information and its relation to Shannon
Information, in Quantum Theory: Reconsideration of Foundations, Ser.
Math. Modeling, 2, V\"axj\"o Univ. Press,  2002.

\end{document}